\documentclass[12pt]{article}
\normalbaselines
\begin{document}
\bibliographystyle{unsrt}
\vskip 2em \begin{center}
 {\LARGE\bf 
Asymptotic quantum estimation theory \par for the displaced thermal states family
\par} \vskip 1.5em 
\large \lineskip .5em
Masahito Hayashi \par
Department of Mathematics, Kyoto University, Kyoto 606-8502, Japan \par 
e-mail address: masahito@kusm.kyoto-u.ac.jp \par
\end{center}
\abstract{Concerning state estimation, we will compare two cases.
In one case we cannot use the quantum correlations between samples.
In the other case, we can use them.
In addition, under the later case, we will propose
a method which simultaneously measures
the complex amplitude and the expected photon number
for the displaced thermal states.}

\section{Introduction}
Quantum estimation is essentially different from
classical estimation regarding 
the following two points.
The first point is that
we cannot simultaneously construct the optimal estimators
corresponding to respective parameters
because of non-commutativity between them.
It has been a serious problem since the beginning of the quantum estimation
\cite{Hel,YL,HolP}.
The second point is that we can reduce the 
estimation error under the assumption that
we can prepare independent and identical samples of
the unknown quantum state.
It was pointed by Nagaoka \cite{Na1,Na2} concerning
the large deviation theory in one-parameter estimation.
The purpose of this paper is to clear the second point
concerning the mean square error (MSE).

Our situation is divided into the following two cases.
In the first case, we estimate the unknown state by independently measuring 
every sample.
In this case, we may decide the $n$-th POVM from
$n-1$ data which have been already given.
In the second case, we estimate the
unknown state by regarding $n$-sample system as a single composite system.
In this case, we may use POVMs which 
are indivisible into every sample system.
In order to construct these POVMs, we need to use 
quantum correlations between every sample.
The former is called the non-quantum correlation case and
the later the quantum correlation case.
When the unknown state is a pure
state,
the errors of both are asymptotically equivalent
in the first order \cite{Haya}.
Concerning the spin 1/2 system, see Hayashi \cite{Hay}.

In this paper, we formulate a general theory for
the asymptotic quantum estimation.
It is applied to the simultaneous estimation
of the expected photon number and the complex amplitude for 
the quantum displaced thermal state.
\section{Asymptotic Estimation Theory}
In this paper, we use a quantum state family ${\cal S}$ parameterized
by finite parameters $\theta^1, \cdots , \theta^d$:
\begin{eqnarray*}
{\cal S}:=\{ \rho_{\theta} \in {\cal S}({\cal H}) | \theta = ( \theta^1 , \ldots , \theta^d ) \in \Theta \subset {\bf R}^d \},
\end{eqnarray*}
where the set ${\cal S}({\cal H})$ denotes the set of densities on ${\cal H}$.
For simplicity, we assume that $\rho_{\theta}$ is nondegenerate.
\subsection{Non-quantum correlation case}
The non-quantum correlation case is formulated as follows.
A pair ${\cal E}_n=( \{ M_k \}_{k=1}^n , \hat{\theta_n})$
is called a {\it recursive estimator} where
$\hat{\theta_n}$ is a function estimating the unknown parameter from
$n$ data, and $\{ M_k \}_{k=1}^n$ is a sequence of POVMs 
$M_1 ,M_2(\omega_1) , \ldots, M_n(\omega_1 , \ldots , \omega_{n-1})$
as follows:
the $n$-th POVM $M_k(\omega_1 , \ldots , \omega_{k-1})$ is determined by $k-1$
data which have been already given.
A sequence $\{ {\cal E}_n \}_{n=1}^{\infty}$ of recursive estimators 
is called a {\it recursive MSE consistent estimator} if 
\begin{eqnarray*}
\lim_{n\to \infty}
\underbrace{\int \int \cdots \int}_{n}
\left\| \hat{\theta_n}(\omega_1, \omega_2 , \ldots , \omega_n ) - \theta 
\right\|^2
P_{\rho_{\theta}}^{{\cal E}_n}
( \,d \omega_1 ,\,d \omega_2, \ldots , \,d\omega_n)
=0,~ \forall \theta \in \Theta,
\end{eqnarray*}
where
\begin{eqnarray*}
 P_{\rho}^{{\cal E}_n}( \,d \omega_1 , \,d \omega_2, \ldots , \,d\omega_n) 
= \mathop{\rm tr}\nolimits \rho M_1 ( \,d \omega_1)
\mathop{\rm tr}\nolimits \rho M_2( \omega_1 )( \,d \omega_2 ) 
\ldots
\mathop{\rm tr}\nolimits 
\rho M_n( \omega_1 , \ldots , \omega_{n-1} )( \,d \omega_n).
\end{eqnarray*}
We define the {\it non-quantum-correlational Cram\'{e}r-Rao type bound} 
$C_{\theta}^{NQC}(G)$ for
a weighted matrix $G$($G$ is a $d \times d$ real positive symmetric
matrix.) as:
\begin{eqnarray*}
C_{\theta}^{NQC}(G):=
\inf \left \{ \left.
\liminf_{n \to \infty}
\mathop{\rm Tr} n G V_{\theta}( {\cal E}_n) \right|
\{ {\cal E}_n \}_{n=1}^{\infty}
\hbox{ is a recursive MSE consistent estimator } \right \},
\end{eqnarray*}
where the MSE matrix $V_{\theta}({\cal E}_n)$ is given by:
\begin{eqnarray*}
V_{\theta}^{i,j}({\cal E}_n)=
\underbrace{\int \cdots \int}_{n}
\left( 
 \hat{\theta_n}^i(\omega_1, \ldots , \omega_n ) - \theta^i 
\right )
\left( 
 \hat{\theta_n}^j(\omega_1, \ldots , \omega_n ) - \theta^j
\right )
P_{\rho_{\theta}}^{{\cal E}_n}( \,d \omega_1 , \ldots , \,d \omega_n).
\end{eqnarray*}
We have the following equation:
\begin{eqnarray*}
C_{\theta}^{NQC}(G)=
\inf \left\{ \left.
\mathop{\rm Tr} G \left( J_{\theta}^{M}\right)^{-1} \right|
M \hbox{ is a POVM on } {\cal H} \right\},
\end{eqnarray*}
where
$J_{\theta}^{M}$ denotes the Fisher information matrix at $\theta$ of 
$\{ \mathop{\rm tr}\nolimits\rho_{\theta} M(\,d \omega) 
| \theta \in \Theta \}$.
It is derived by Jensen's inequality \cite{HM}.
Under some regular condition,
we show that there exists a recursive MSE consistent estimator 
$\{{\cal E}_n \}_{n=1}^{\infty} $ such that \cite{HM}:
\begin{eqnarray*}
n \mathop{\rm Tr} G V_{\theta}({\cal E}_n) \to C_{\theta}^{NQC}(G) 
\hbox{ as }n \to \infty 
,~ \forall \theta \in \Theta.
\end{eqnarray*}
According Holevo\cite{HolP}, we have
$J_{\theta}^M \le \tilde{J}_{\theta} $,
where $\tilde{J}_{\theta}$ is the {\it RLD Fisher information matrix}
defined as:
$
\tilde{J}_{\theta;i,j} := 
\mathop{\rm tr}\nolimits (\tilde{L}_{\theta;i})^* \rho_{\theta}\tilde{L}_{\theta;j} , ~
\tilde{L}_{\theta;i} :=
(\rho_{\theta})^{-1} \frac{\partial \rho_{\theta}}{\partial \theta^i}.
$
Therefore we have the following inequality
\begin{eqnarray*}
C_{\theta}^{NQC}(G) \ge C_{\theta}^R(G),
\end{eqnarray*}
where 
\begin{eqnarray*}
C_{\theta}^R(G) &:=&
\inf\left\{
\mathop{\rm tr}\nolimits GV | V \hbox{ is a } d \times d \hbox{ real symmetric matrix }
V \ge \tilde{J}_{\theta}^{-1} \right\} \\
&=& \mathop{\rm Tr} G \mathop{\rm Re} \tilde{J}_{\theta}^{-1} + 
\mathop{\rm Tr} \left | \sqrt{G} \mathop{\rm Im} \tilde{J}_{\theta}^{-1} \sqrt{G} \right |.
\end{eqnarray*}

\subsection{Quantum correlation case}
Next, we formulate the quantum correlation case.
For this purpose, we consider a quantum counterpart of independent and
identically distributed condition.
If ${\cal H}_1, \ldots , {\cal H}_n$ are $n$ Hilbert spaces which correspond 
to the physical systems,
then their composite system is represented by the tensor Hilbert space.
\begin{eqnarray*}
{\cal H}^{(n)}:=
{\cal H}_1 \otimes \cdots \otimes {\cal H}_n .
\end{eqnarray*}
Thus, a state on the composite system is denoted by a density operator 
$\rho^{(n)}$ on ${\cal H}^{(n)}$.
In particular if $n$ element systems (${\cal H}_1, \ldots , {\cal H}_n$) 
of the composite system
${\cal H}^{(n)}$ are independent of each other,
there exists a density $\rho_k$ on ${\cal H}_k$ such that
\begin{eqnarray*}
\rho^{(n)}= \rho_1 \otimes \cdots \otimes \rho_n , 
\hbox{ on }{\cal H}^{(n)}.
\end{eqnarray*}
The condition:
\begin{eqnarray*}
{\cal H}_1 = \cdots = {\cal H}_n={\cal H},~
\rho_1= \cdots = \rho_n =\rho 
\end{eqnarray*}
corresponds to the independent and identically distributed condition
in the classical case.
Therefore, we consider the parameter estimation
problem for the family 
$\{ \rho_{\theta}^{(n)}  \hbox{ on } {\cal H} |
\theta \in \Theta \}$
which is called the {\it $n$-i.i.d. extended family}.

In this case, we use a sequence $\{ M^n \}_{n=1}^{\infty}$
of POVMs where $M^n$ is a POVM on ${\cal H}$ whose measurable set
is ${\bf R}^d$ as an estimator.
A sequence $\{  M^n \}_{n=1}^{\infty}$
is called an {\it MSE consistent estimator}
if $\{  M^n \}_{n=1}^{\infty}$ satisfies 
\begin{eqnarray*}
\lim_{n \to \infty}
\int_{{\bf R}} 
\| \hat{\theta}- \theta \|^2 \mathop{\rm tr}\nolimits \rho^{(n)}_{\theta} M^n(\,d \hat{\theta})
=0 ,~ \forall \theta \in \Theta.
\end{eqnarray*}
A recursive MSE consistent estimator can be regarded as
an MSE consistent estimator because a recursive estimator 
${\cal E}_n= ( \{ M_k \}_{k=1}^n , \hat{\theta}_n)$ 
is regarded as a POVM $M({\cal E}_n)$
as follows:
\begin{eqnarray*}
M({\cal E}_n)(B):=
\int_{\hat{\theta}_n^{-1}(B)} \bigotimes_{k=1}^n M_k (\omega_1 ,
\ldots, \omega_{k-1})(\,d \omega_k) ,~
\forall B \subset {\bf R}^d \hbox{ on } {\cal H}^{(n)}.
\end{eqnarray*}
We define the {\it quantum-correlational Cram\'{e}r-Rao type bound} 
$C_{\theta}^{QC}(G)$
for a weighted matrix $G$ as:
\begin{eqnarray*}
C_{\theta}^{QC}(G):=
\inf\left\{ \left.
\liminf_{n \to \infty} n \mathop{\rm Tr} G V_{\theta}(M^n) \right|
\{ M^n \}_{n=1}^{\infty} \hbox{ is an MSE consistent estimator } \right\}
\end{eqnarray*}
where the MSE matrix $ V_{\theta}(M^n)$ is given by:
\begin{eqnarray*}
 V_{\theta}^{i,j}(M^n)
=
\int_{{ \bf R}^d} (\hat{\theta}^i - \theta^i )(\hat{\theta}^j - \theta^j )
\mathop{\rm tr}\nolimits \rho_{\theta}^{(n)} M^n (\,d \hat{\theta}).
\end{eqnarray*}
We have the following equation
\begin{eqnarray*}
C_{\theta}^{QC}(G) = \liminf_{n \to \infty} n C_{\theta}^n(G) ,
\end{eqnarray*}
where $C_{\theta}^n(G) $ denotes 
the non-quantum-correlational Cram\'{e}r-Rao type bound 
for the $n$-i.i.d. extended family \cite{HM}.
From the definition of the $n$-i.i.d. extended family,
we have $ C_{\theta}^{NQC}(G) \ge n C_{\theta}^n(G) $.
Therefore, we have the first inequality of (\ref{bbb}).
\begin{eqnarray}
C_{\theta}^{NQC}(G) \ge C_{\theta}^{QC}(G) \ge C_{\theta}^R(G). \label{bbb}
\end{eqnarray}
It shows the second inequality of (\ref{bbb}) that 
$J_{\theta}^{M^n} \le n \tilde{J}_{\theta} $
for any POVM $M^n$ on ${\cal H}^{(n)}$.
Therefore the difference between 
$C_{\theta}^{NQC}(G)$ and $C_{\theta}^{QC}(G)$ means the difference of the 
quantum correlation case from the non-quantum correlation case.
Under some regular condition,
we can show that there exists an MSE consistent estimator 
$\{M^n \}_{n=1}^{\infty} $ such that \cite{HM}:
\begin{eqnarray*}
n \mathop{\rm Tr} G V_{\theta}( M^n) \to C_{\theta}^{QC}(G) 
\hbox{ as }n \to \infty 
,~ \forall \theta \in \Theta.
\end{eqnarray*}
\section{Quantum displaced thermal states family}
Now we consider the estimation for the the complex amplitude 
$\zeta$ and expected photon number $N$ 
for the quantum displaced thermal states family defined as:
\begin{eqnarray*}
{\cal S}:= 
\left\{ \left. \rho_{\zeta,N} := \frac{1}{\pi N}
\int_{{\bf C}} \exp \left( - \frac{| \zeta - \alpha | ^2}{N} \right) | \alpha \rangle 
\langle \alpha | \,d^2 \alpha \right| \zeta \in {\bf C}, N \,>0 \right\}.
\end{eqnarray*}
\subsection{Estimation of complex amplitude $\zeta$} 
In the case of that photon number $N$ is known,
we estimate the tow unknown parameters
$\zeta= (\theta^1 +  i \theta^2)/\sqrt{2}$.
This estimation problem is investigated by 
Yuen, Lax and Holevo \cite{YL,HolP}.
In this case they calculated the inverse $\tilde{J}_{\theta}^{-1}$
of the RLD Fisher information matrix as:
\begin{eqnarray*}
\tilde{J}_{\theta}^{-1} =
\left( 
\begin{array}{cc}
N+\frac{1}{2}  & \frac{i}{2} \\
- \frac{i}{2} & N+\frac{1}{2}
\end{array}
\right).
\end{eqnarray*}
They calculated the non-quantum-correlational Cram\'{e}r-Rao type bound 
$C_{\theta}^{NQC}(G)$
as follows:
\begin{eqnarray}
C_{\theta}^{NQC}(G)= C_{\theta}^R(G) =2\left (N+\frac{1}{2} \right) g_1 +
\sqrt{g_1^2 - g_2 ^2 - g_3^2}
, \label{saku}
\end{eqnarray}
where the weighted matrix $G$ is parameterized as:
\begin{eqnarray*}
G= 
\left( 
\begin{array}{cc}
g_1 + g_2 & g_3 \\
g_3 & g_1 - g_2 
\end{array}
\right).
\end{eqnarray*}
From (\ref{bbb}) and (\ref{saku}), we have
the following equations.
\begin{eqnarray}
C_{\theta}^{NQC}(G) = C_{\theta}^{QC}(G) = C_{\theta}^R(G) . \label{111}
\end{eqnarray}
In this case, 
the optimal estimator is the squeezed heterodyne.
\subsection{Simultaneous estimation of complex amplitude $\zeta$ 
and expected photon number $N$} 
Next we consider the case of that both of the expected photon number $N$ and
the complex amplitude $\zeta$ are unknown.
In this case, 
we estimate three unknown parameters $\theta^1, \theta^2$ and $\theta^3=N$.
The first equation of (\ref{111}) isn't held.
Therefore, the squeezed heterodyne isn't optimal.
The inverse $\tilde{J}_{\theta}^{-1}$ 
of the RLD Fisher information matrix is calculates as:
\begin{eqnarray*}
\tilde{J}_{\theta}^{-1} =
\left( 
\begin{array}{ccc}
N+\frac{1}{2}  & \frac{i}{2} &0\\
- \frac{i}{2} & N+\frac{1}{2} & 0 \\
0 & 0& N(N+1)
\end{array}
\right).
\end{eqnarray*}
Therefore
we can calculated $C_{\theta}^R(G)$ as:
\begin{eqnarray*}
C_{\theta}^R(G) =
g_0 N(N+1) + 2\left (N+\frac{1}{2} \right) g_1 +
\sqrt{g_1^2 - g_2 ^2 - g_3^2}
, 
\end{eqnarray*}
if the weighted matrix $G$ can be parameterized as:
\begin{eqnarray}
G= 
\left( 
\begin{array}{ccc}
g_1 + g_2 & g_3 &0 \\
g_3 & g_1 - g_2 & 0 \\
0& 0 & g_0
\end{array}
\right). \label{122}
\end{eqnarray}
If the weighted matrix $G$ can be parameterized as (\ref{122}),
we obtain the following equations:
\begin{eqnarray}
C_{\theta}^{NQC} (G) \,> C_{\theta}^{QC}(G) = C_{\theta}^R(G) .
 \label{112}
\end{eqnarray}
A proof for $C_{\theta}^{NQC} (G) \,> C_{\theta}^{QC}(G)$ is omitted.
The inequality $C_{\theta}^{NQC} (G) \,> C_{\theta}^{QC}(G)$ means 
that we cannot the simultaneous measurement
of the photon number counting and heterodyne for a single sample.
\subsection{Construction of an MSE consistent estimator 
$\{ M^n \}_{n=1}^{\infty}$ attaining $ C_{\theta}^R(I) $ }
Now, for a weighted matrix $I$,
we construct an MSE consistent estimator $\{ M^n \}_{n=1}^{\infty}$ such that 
\begin{eqnarray*}
\lim_{n \to \infty} 
n \mathop{\rm tr}\nolimits V_{\theta}(M^n)=
C_{\theta}^R(I) ,~ \forall \theta \in \Theta .
\end{eqnarray*}
It is sufficient for $C_{\theta}^{QC}(I) = C_{\theta}^R(I) $ 
to construct such an MSE consistent estimator.

Every POVM $M^n$ is constructed in the following step:

(1) Evolve the unknown state  
$\underbrace{\rho_{\zeta,N} \otimes \cdots \otimes \rho_{\zeta,N}}_{n}$
as:
\begin{eqnarray*}
\underbrace{\rho_{\zeta,N} \otimes \cdots \otimes \rho_{\zeta,N}}_{n}
\to 
U_n \underbrace{\rho_{\zeta,N} \otimes \cdots \otimes  \rho_{\zeta,N}}_{n}U_n^*
= \rho_{\sqrt{n} \zeta,N} \otimes 
\underbrace{\rho_{0,N} \otimes \cdots \otimes \rho_{0,N}}_{n-1}
\hbox{ on } {\cal H}^{(n)},
\end{eqnarray*}
where
\begin{eqnarray*}
U_n &=&
\exp \phi_{n-1} (a_n^* a_1 - a_1^* a_n )
\cdots
\exp \phi_2 ( a_3^* a_1 - a_1^* a_3 )
\exp \phi_1 ( a_2^* a_1 - a_1^* a_2 )
 \hbox{ on } {\cal H}^{(n)}\\
\phi_i &=& \arctan \frac{1}{\sqrt{i}} , \quad i=1 ,2 , \ldots, n-1.
\end{eqnarray*}
$a_i$ denotes the annihilation operator on ${\cal H}_i$.

(2) Measure the first sample $\rho_{\sqrt{n} \zeta,N}$ by the heterodyne,
then we get the estimate $\hat{\zeta}$ of the complex amplitude.

(3) Measure the others by the photon counting, then we obtain
$n-1$ data which obey the probability distribution $P^N(k)$:
\begin{eqnarray*}
P^N(k)=
\frac{1}{N+1}\left( \frac{N}{N+1} \right)^k, \quad
k=0,1 \ldots .
\end{eqnarray*}

\indent(4) We obtain the estimate $\hat{N}$ of the expected 
photon number $N$ by the maximum likelihood estimator of
the probability distribution $ P^N(k_1) , \ldots ,P^N(k_{n-1})$.

\section{Conclusion}
We formulate an asymptotic quantum estimation theory.
This theory is applied to the simultaneous measurement
of the photon number counting and the heterodyne for displaced thermal states.
It is a future study to realize the MSE consistent estimator proposed 
in this paper in an actual physical system.

\section{Acknowledgments}
This work was supported by the Research Fellowships of the Japan Society for
the Promotion of Science for Young Scientists No. 9404.
The author would like to thank Dr. K. Matsumoto for
useful discussions of these topics.


\begin{thebibliography}{99}
\bibitem{Hel} C. W. Helstrom.
``Quantum Detection and Estimation Theory,''
Academic Press, New York (1976).

\bibitem{YL}
H. P. Yuen and M. Lax,
Multiple-parameter quantum estimation and measurement of 
nonselfadjoint observables,
{\it IEEE trans. Inform. Theory,} {\bf IT 19}, 740 (1973).

\bibitem{HolP} A. S. Holevo.
``Probabilistic and Statistical Aspects of Quantum Theory,''
North\_Holland, Amsterdam (1982).

\bibitem{Na1} H. Nagaoka,
On the relation Kullback divergence and Fisher information
-from classical systems to quantum systems-, {\it in}:
``Proc. Society Information Theory and its Applications in Japan''
(1992)(in Japanese).

\bibitem{Na2} H. Nagaoka,
Two quantum analogues of the large deviation Cra\'{m}r-Rao inequality,
{\it in}:
``Proc. 1994 IEEE Int. Symp. on Information Theory'' p.118
(1994).

\bibitem{Haya} M. Hayashi,
Asymptotic estimation theory for a finite-dimensional
pure state model,
{\it J. Phys. A: Math. Gen.} {\bf 31} 4633 (1998).

\bibitem{Hay}  M. Hayashi,
Asymptotic Quantum Parameter Estimation in Spin 1/2 System,
LANL e-print quant-ph/9710040 (1997).

\bibitem{HM} M. Hayashi and K. Matsumoto,
Quantum mechanics as a statistical model which
allows a free choice of measurements
{\it in}:
``Large Deviation and Statistical Inference''
RIMS koukyuuroku No. 1055 RIMS, Kyoto (1998)
(in Japanese).
\end{thebibliography}
\end{document}